# THE SELFORGANIZATION OF VACUUM, PHASE TRANSITIONS AND THE COSMOLOGICAL CONSTANT

V. Burdyuzha[1], O. Lalakulich[2], Yu. Ponomarev[1], G. Vereshkov[2]

[1] Astro Space Center of Lebedev Physical Institute of Russian Academy of Sciences, Profsoyuznaya 84/32, 117810 Moscow, Russia

[2] Rostov State University, Stachki str. 194, 344104 Rostov on Don, Russia,

**Abstract**

The problem of the physical nature and the cosmological genesis of $\Lambda$-term is discussed. This problem can't be solved in terms of the current quantum field theory which operates with Higgs and non-perturbative vacuum condensates and takes into account the changes of these condensates during relativistic phase transitions. The problem can't be completely solved also in terms of the conventional global quantum theory: Wheeler-DeWitt quantum geometrodynamics does not describe the evolution of the Universe in time (RPT in particular). We have investigated this problem in the context of energies density of different vacuum subsystems characteristic scales of which pervaid all energetic scale of the Universe. At first the phemenological solution of $\Lambda$-term problem and then the hypothesis about the possible structure of a new global quantum theory are proposed. The main feature of this theory is the irreversible evolution of geometry and vacuum condensates in time in the regime of their selforganization. The transformation of the cosmological constant in dynamical variable is inevitably.

The cosmological constant problem is one of intriguing problem of modern physics and cosmology. The suggestions for its solution have attracted a lot of attention [1]. The idea of compensation of a initial vacuum energy by vacuum condensates of quantum fields with cooling of cosmological plasma during relativistic phase transitions is discussed (the short version of this article using Zel'dovich's approximation was be published in [2]). The list of other adjustment mechanisms for cosmological constant can also find in [1]. The terms of cosmological constant ($\Lambda$-term) and vacuum energy are used practically synonymously in modern cosmology.

Today the strictly established fact is the physical vacuum is a complex heterogenic system of classic and quantum fields consisting of three subsystems: 1) zeroth weakly correlated vibrations of quantum fields; 2) zeroth strongly correlated vibrations of quantum fields producing nonperturbative vacuum condensates; 3) quasiclassic (quasihomogeneous and quasistationary) fields usual named Higgs condensates. All these subsystems are included in Standard Model (SM). The existence of zeroth weakly correlated vibrations has the experimental confirmation: anomalous magnetic moment of electron, Lamb's shift, Casimir effect, radiative corrections. The values of non-perturbative condensates firstly have introduced in [3] are usually established in physics of vector mesons. The question of Higgs condensates existence will be decided after detection of Higgs bosons. Here we are discussing the cosmological constant problem in context of density energies of different vacuum subsystems. The problem is that each vacuum subsystem has a huge density energy, however the total value of vacuum energy in the Universe today is near zero as observational data confirm probably [4]. Thus the phenomenon of selforganization of vacuum is evident although the mechanism of selforganization of nonperturbative condensates does not understand till now well (the possible models of "nullification" of vacuum energy have been discussed also in [5-6]).

Here selforganization of vacuum is the ability of a system to react on outer condition such way to conserve itself local stability and to evolve subsequently. Other aspect is that vacuum is a strongly nonlinear system the intensity interactions in which depends on outer conditions. The third aspect is the value $\Lambda \approx 0$ sure to be cosmologically preferable. The Universe with a large negative $\Lambda$-term never become macroscopic, if the value of $\Lambda$-term is large and positive the production of complex nuclear, chemical, biological and cosmogonical structures is impossible. Our real Universe with the observed structure hierarchy can exist when $\Lambda \sim 0$ only.

The energy density of vacuum in General Relativity is described by constant $\Lambda$-term when the interactions of the vacuum subsystems with matter is negligible:

$$R_{\mu\nu} - \frac{1}{2}g_{\mu\nu}R = 8\pi G T_{\mu\nu} + \Lambda g_{\mu\nu}. \tag{1}$$

The problem is to calculate this constant in the modern and previous epochs. Today we can more exactly define the physical meaning of $\Lambda$-term which must contain the energy-momention tensor (EMT) of gravitational vacuum $T_{\mu\nu(g)}$ and EMT of quantum fields $T_{\mu\nu(QF)}$:

$$R_{\mu\nu} - \frac{1}{2}g_{\mu\nu}R = æ(T_{\mu\nu(g)} + <T_{\mu\nu(QF)}>) = æ(g_{\mu\nu}\Lambda_g + g_{\mu\nu}\Lambda_{QF}). \tag{2}$$

Here $æ = (10^{19}\ Gev)^{-2}$ is the gravitational constant in the system of units where $\hbar = c = 1$; $<T_{\mu\nu(QF)}>$ is EMT of quantum field averaged on some martix of density, which contains the information about state of plasma and vacuum of elementary particles. For $\mid R_\mu^\nu \mid \ll æ^{-1}$ the averaged EMT of quantum fields is:

$$<T_{\mu\nu(QF)}> = <0\mid T_{\mu\nu(QF)}\mid 0> = g_{\mu\nu}\Lambda_{QF};$$

$$T_{\mu\nu(g)} = g_{\mu\nu}\Lambda_g.$$

Here $\Lambda_g$ is the second fundamental constant of the gravitation theory accounting a gravitational vacuum condensate. Other words $\Lambda$-term must contain two items:

$$\Lambda = \Lambda_g + \Lambda_{QF} \tag{3}$$

which have practically exactly compensated each other since the observable value of $\Lambda$-term (the cosmological constant) is near zero [4]. Many theorists find that $\Lambda$-term must be calculated in a unified theory of all interactions and the separation on two items is artificial (naturally it is so). But the subject of our research is a heterogenic system (geometry + vacuum + fields of matter) and for a arbitrary state of this system is not possible to extract the vacuum energy as the separate item in $<T_{\mu\nu(QF)}>$. The constant $\Lambda_{QF}$ can arise as a physical magnitude if two conditions are carried out: 1) a vacuum subsystem after relaxation must reach the equilibrium state with plasma of elementary particles; 2) the temperature and density of plasma must be small in comparison with critical values of magnitudes which characterize the point of a relativistic phase transition (RPT) (more detail see [7]). For illustration we shall use the simplest chain of RPT which may take place during initial evolution in our Universe:

$$\text{P} \underset{10^{19}Gev}{\Longrightarrow} D_4 \times [SU(5)]_{SUSY} \underset{10^{16}Gev}{\Longrightarrow} D_4 \times [U(1) \times SU(2) \times$$

$$\times SU(3)]_{SUSY} \underset{10^5 - 10^{10}Gev}{\Longrightarrow} D_4 \times U(1) \times SU(2) \times SU(3)$$

$$\underset{10^2 Gev}{\Longrightarrow} D_4 \times U(1) \times SU(3) \underset{150 Mev}{\Longrightarrow} D_4 \times U(1)$$

Of course whole chain is our proposal but one can be sure only in two last transitions: electroweak (EW PT) and quark-hadron PT. EW PT occurs at temperature about $10^2$ GeV and is accompanied by appearence of a Higgs condensate decreasing vacuum energy. In the interval of temperatures $150\ Mev < T < 100\ Gev$ the vacuum in the Universe was in the state of spontaneously breaking $SU(2)_L$ symmetry (for these temperatures the quark-gluon subsystem was in the state of de-confinement that is the quark-gluon vacuum condensate was absent). The value of Higgs condensate is negative and it have been calculated many times in different models. We use an expression:

$$\Lambda_{SM} = -\frac{m_H^2 m_w^2}{2g^2} - \frac{1}{128\pi^2}(m_H^4 + 3m_z^4 + 6m_w^4 - 12m_t^4). \quad (4)$$

Here the first term is the energy density of a quasiclassical Higgs condensate. The second term is the change of energy density of bosonic and fermionic fields by the quasiclassical field of a condensate( that is the polarization of vacuum by quantum fields takes into account). Excepting $t$-quark others fermions are very light and they involve a negligible small contribution in formula (4). Boson contributions are negative but fermion ones are positive. The numerical values of all constants except for Higgs boson mass are known from experiments (see [8]). The limitations on a Higgs boson mass can be found from the condition of vacuum stability:

$$x^2 + x(\frac{1}{2a} - \frac{4ab}{9}) - \frac{2b}{3} > 0,$$

$$x < \frac{1}{a} + \frac{4ab}{9}, \quad (5)$$

here $x = m_H^2/m_w^2; a = 3g^2/128\pi^2; b = \frac{12m_t^4 - 3m_z^4 - 6m_w^4}{m^4}, g^2 = 0.43$ is the gauge constant $SU_L(2)$ group; $m_w = 80\ Gev, m_z = 91\ Gev, m_t = 175\ Gev$. Inequalities (5) give the interval of possible values of still no experimentally detected Higgs boson $36\ GeV < m_H < 2480\ GeV$ that does not contradict modern experimental restrictions $m_H > 75\ GeV$. Substituting these values of $m_H$ into (4) one can see that the mutual compensation of positive and negative contributions in vacuum density energy in SM is prohibited by the condition of stability. The last conclusion has a general character. It is an important moment of our consideration. For decreasing of any symmetry during RPT the vacuum energy must decrease. So in order to have $\Lambda \approx 0$ one have to introduce ad hoc the initial positive $\Lambda$-term. Then the decreasing of vacuum energy during RPT can be considered as the compensation of an initial positive value. It seems to be the phenomenological solution of the $\Lambda$-term problem.

The next question arises immediately. Can zeroth value of $\Lambda$-term be obtained as the consequence of the inner structure of a theory? The answer was searching in terms of SUSY theories. The idea is based on the main feature of such theories: the general contributions from Higgs condensates are nearly compensated. The residual part must be compensated by radiative corrections. This has been made in multidimentional superstring model after the special compactification in one-loop approximation. The possibility of such coordination is prompted by mathematical formalism of strongly nonlinear theory in which the state with $\Lambda = 0$ has the status of a special branch of nonlinear equations. Quite evidently that the coordination of vacuum subsystem states was realized during cosmological evolution that is here we have all indicators of vacuum selforganization. It is pertinently to recall the anthropic principle (probably the selforganization of vacuum has provided the life of organic type in the Universe). We want to stop also more detail on QCD nonperturbative vacuum since the extrapolation of QCD ideas to more deep structure levels of matter [9] and to quantum gravity scales [10] is almost inevitably. A nonperturbative quark-gluon condensate as the element of the theory is included in SM . Without representations about this condensate the confinement phenomena of quarks and gluons is not possible to understand. The investigation of QCD equations has shown that vacuum possesses of confinement properties if vacuum correlatories of quark- gluon fields is not zero:

$$< 0 \mid G_{\mu\nu} G^\mu \mid 0 > \ > 0; \quad < 0 \mid \bar{q}q \mid 0 > \ < 0 \qquad (6)$$

(here for simplisity we are limiting the discussion of quantum correlatories which are quadratic from fields but certainly in the nonperturbative vacuum correlatories of any order are not zero). In the perturbative vacuum these values after renormalization equal zero. Inequalities (6) have the status of rigorous theoretical results however they say nothingcondensates has been found [3]:

$$< 0 \mid \frac{\alpha_s}{\pi} G_{\mu\nu} G^{\mu\nu} \mid 0 > = (360 \pm 20 \ MeV)^4 \approx 27 \lambda_{QCD}^4 \tag{7}$$

$$< 0 \mid \bar{u}u \mid 0 > \approx < 0 \mid \bar{d}d \mid 0 > \approx < 0 \mid \bar{s}s \mid > \approx (-225 \pm 25 \ MeV)^3 \approx -2.8 \lambda_{QCD}^3$$

and then the density energy of nonperturbative QCD vacuum is:

$$\epsilon_{vac} = -\frac{9}{32} < 0 \mid \frac{\alpha_s}{\pi} G_{\mu\nu} G^{\mu\nu} \mid 0 > + \frac{1}{4} < 0 \mid m_u \bar{u}u \mid 0 > +$$

$$+ < 0 \mid m_d \bar{d}d \mid 0 > + < 0 \mid m_s \bar{s}s \mid 0 > = -8.2 \lambda_{QCD}^4. \tag{8}$$

Here $m_u = 4.2 \ Mev; m_d = 7.5 \ Mev; m_s = 150 \ Mev$ are masses of light quarks satisfacting to the condition $m_q \leq \lambda_{QCD}; \lambda_{QCD} = 160 \ Mev$. What is the physical nature of nonperturbative fluctuations forming a quark-gluon condensate? Probably the nonperturbative vacuum is a bose-condensate of dions and antidions. The sum charges of vacuum and averaged on large distances gluon (chromoelectrical and chromomagnetic) fields in vacuum equal of course zero however fluctuations of these fields in scales of the correlated lenght of a dion condensate are not zero. The average values of square of fluctuating gluon fields is the basic characteristics of nonperturbative QCD-vacuum. Fluctuations of quark fields are probably induced by fluctuations of gluon fields. This follows from the relation:

$$< 0 \mid \bar{q}q \mid 0 > = -\frac{1}{12 \mu_q} < 0 \mid \frac{\alpha_s}{\pi} G_{\mu\nu} G^{\mu\nu} \mid 0 >, \tag{9}$$

here

$$\mu_q = \begin{cases} \lambda_{QCD} : q = u, d, s & m_q \leq \lambda_{QCD} \\ m_q : q = c, b, t & m_q \gg \lambda_{QCD} \end{cases}.$$

Thus for $T < \lambda_{QCD} = 160 \ Mev$ a nonperturbative quark-gluon vacuum is the state of a dion condensate with negative energy density (the classic prototype of dions is nonlinear solutions of Yang-Mills equations similar to solitons, instantons, monopoles). That is today vacuum in the Universe has the confinement phase and the modern value of $\Lambda$-term can be calculated using formula:

$$\Lambda_{QF} = \Lambda_{SM} + \epsilon_{vac}. \tag{10}$$

Here $\epsilon_{vac}$ is the energy density of quark-gluon vacuum (see (8)) and $\Lambda_{SM}$ is the constant taking phenomendogicaly into account all vacuum structures on energy scales more than $\lambda_{QCD}$. It can be easily understood that only $\Lambda_{SM}$ coming from Higgs vacuum can be compensated by SUSY mechanism. The similar mechanism for $\epsilon_{vac}$ coming from non-perturbative vacuum is absent until the problem of dimensional transmutation will not be solved. Finally our conclusion is the problem of $\Lambda$-term can't be solved in terms of the current field theory.

It is worthwhile to say some words about calculations of today value of $\Lambda$-term which were carried out recently in [2] using Zeldovich's approximation. The vacuum condensates (Higgs and nonperturbative one) in the modern quantum theory are macroscopic mediums with quasiclassical properties. The periodic collective motions in these mediums are perceived as pseudogoldstone bosons. For temperature of chiral symmetry breaking ($T_c \sim 150 \, Mev$) the main contribution in periodic collective motions of a nonperturbative vacuum quark-gluon condensate introduces $\pi$-mesons as the lightest pseudogoldstone particles. That is here in essence the spectrum of excitations reflects the properties of a ground state. Ya.Zeldovich [11] attempts to account for a nonzero vacuum energy density of the Universe in terms of quantum fluctuations (the gravitational force between particles in the vacuum fluctuations as a higher-order effect) inserting in the finded them formula $\Lambda = 8\pi G^2 m^6 \hbar^{-4}$ the mass of proton or electron. Calculations have shown that the agreement of the result with the observed value is not good. Kardashev [12] had proposed to modify Zeldovich's formula and use mass of pions:

$$\Lambda = 8\pi G^2 m_\pi^6 h^{-4}. \tag{11}$$

(note, here we have $h$ instead of $\hbar$). . Remarkably that the calculated values of $\Lambda$-term using Zeldovich's formula gives $\Omega_\Lambda = 0.7$ if $H_o = 72.5 \, (km/s)/Mps$ (here $H_o$ is the Hubble constant) and $\Omega_\Lambda = 0.8$ if $H_o = 67.8 \, (km/s)/Mps$ (see the table in [2]). The mistic agreement of formula (11) with the observable value nevertheless gives no physical explanation of $\Lambda$-term problem and our conclusion about the necessity to go beyond the current field theory does not change.But it may be the first approximation to our understanding and the calculation of today value of the cosmological constant. The next step can be made by quantum geometrodynamics (QGD).

Initially two approaches have been proposed. The first one is Hawking's idea to introduce in the theory (besides the usual fields describing both vacuum and particles) some special fields which concern to vacuum only. Such special fields were called 3-forms, 4-forms and so on. As S.Hawking has shown [13] that the more probable state of the Universe was the state with $\Lambda_{eff} = 0$ since

$$P(\Lambda_{eff}) \sim exp\,(\frac{3\pi}{\ae^2 \Lambda_{eff}}). \tag{12}$$

here $\Lambda_{eff}$ is the sum of usually discussed $\Lambda$-term and of the contribution from 3-forms. The problem is that is the nature of "formes" and how they can be experimentally observed in a local experiment (beside the influence on $\Lambda$-term). A more deep step in the investigation of $\Lambda$-term problem was made by S.Coleman [14] who took into attention the realistic effect of microscopic quantum fluctuations of space-time topology at the Planck scale (worm holes). In this approach the more probable the state of the Universe has had a more sharp peak than S.Hawking's distribution :

$$P(\Lambda_{eff}) \sim exp\,(exp\,\frac{3\pi}{\ae^2 \Lambda_{eff}}) \tag{13}$$

Here $\Lambda_{eff} = \Lambda_{QF} + \Lambda_{WH}$, where $\Lambda_{WH}$ is the contribution of worm holes. This approach allowes to give the unified conception of vacuum. Recall, the QCD vacuum also is a system of quantum topological fluctuations.

Thus these are two limiting points on the energetic scale of the Universe: the first point is $\Lambda_{QCD} = 150\ MeV$ which are experimental fact described in experimentally tested theories; the second point is fluctuations at the Planck scale which are the direct consequence of Quantum Gravity. Both types of fluctuations have a geometrical origin. We propose the quantum topological fluctuations can exist at other intermediate scales. This idea is realized in preon theories of elementary particles where we have the hierarchy of non-perturbative condensates instead of Higgs condensates. This approach seems to be favourable because it gives the unified picture of vacuum. (recall the Higgs bosons are still experimentally undetected and Higgs conception of vacuum is not confirmed).

Certainly neither Hawking's approach not Coleman's approach do not solve the $\Lambda$-term problem because their results were obtained in terms of Wheeler-De Witt QGD which does not describe the quantum evolution of the Universe. In the real Universe the energy of vacuum has changed in time in the processes of the RPT (this was inevitably).

To solve this problem, a new version of QGD have to be formulated. The new theory must theoretically describe the evolution of the Universe wave function in time. The dynamical processes in vacuum and elementary particle plasma which influence on wave function evolution should be taking into account. As we think at least three steps should be made.

The first step is to change the status of $\Lambda$-term from the cosmological constant to the dynamical variable. In classical theory it was described by Weinberg in [6] who has rewritten the Einstein's equations in the special gauge in the form no containing $\Lambda$-term. In his theory $\Lambda$-term is an integral of motion. Having fixed $\Lambda$-term one finds solutions of equations.

We propose at the second step one should do the same in quantum theory. Here, in distinguishing from classical theory the integral of motion can't have an arbitrary value. The spectrum of the allowed values is fixed by the eigenvalues of the superhamiltonian. This spectrum can be discrete or continuous but from our point of veiw near the small values of $n$ it is discrete. This superhamiltonian wil describe interactions between topological fluctuations of different scales. Any changes at one scale lead to rebuilding of vacuum condensates at other scales. That is we refer to as quantum selforganization of vacuum. The usual quantum theory is the reversible theory (the quantum transitions between the levels with different $\Lambda$-term can go forward and backward and there is no preferable value of $\Lambda$-term). So at the third step it is worth to recall the R.Penrose suggestion that quantum evolution is to be irreversible. This can be realized, for example, by suppression of quantum transitions with $\mid \Lambda \mid$ increasing.

Finally in short way the result is expressed by the formula:

$$\Lambda = \Lambda_{QF} + \Lambda_{WH} + \Lambda_G. \tag{14}$$

Here $\Lambda_{QF}$ is formed by the zeroth vibrations of quantum fields and by non-perturbative condensates; $\Lambda_{WH}$ is formed by worm holes; the evolution of a gravitation vacuum condensate (GVC) forms $\Lambda_G \equiv \dfrac{9\pi^2}{2æ^2}\lambda_n$ where $\lambda_n$ defines the spectrum of GVC possible states. The general for all items in this formula is that they were created during evolution of the Universe.

The value $\Lambda \approx 0$ (but no $\Lambda = 0$) can be explained in the following way. According to mentioned above hypothesis about the discrete spectrum the line with $\Lambda = 0$ must absent but anyway there is a line with $\Lambda$ nearest to zero. During Universe evolution the series of quantum transitions lead to this state as the final state. The inverse transitions, according to R.Penrose, must be suppressed. This is a process of vacuum selforganization. A strategy for a vanishing cosmological constant suggested recently by S.Adler [1] has also included understanding of changes in a vacuum sector in the presence of scale invariance breaking.